\documentstyle[12pt]{article}
\newcommand{\beq}{\begin{equation}}
\newcommand{\eeq}{\end{equation}}
\newcommand{\bea}{\begin{eqnarray}}
\newcommand{\eea}{\end{eqnarray}}

\begin{document}

\begin{flushleft}
{\it Yukawa Institute Kyoto}
\end{flushleft} 

\begin{flushright}
YITP-95-21 \\
January, 1996 \\
hep-ph/9601339 \\
\end{flushright} 
\vspace{1.0cm}

\begin{center}
\Large\bf{Effect of Quark Scatter on Baryogenesis\\
by Electroweak Strings}
\vskip 1.0cm

\large{Michiyasu NAGASAWA}
\footnote{Electronic address: nagasawa@yukawa.kyoto-u.ac.jp}
\vskip 0.2cm

\large\sl{Yukawa Institute for Theoretical Physics, \\
Kyoto University, Kyoto 606-01, Japan}
\end{center}

\begin{abstract}
The amount of baryon asymmetry generated by the electroweak strings
at a rough estimate is smaller than that generated by the bubble expansion
since the collapsing strings cannot cover the whole volume of the universe.
However, the interaction cross section of the strings with particles
can be larger than the geometrical area of the order of the string
thickness. The scattering cross section of quarks from the electroweak
strings is calculated using Dirac equation in the string background.
It it proportional to the reverse of the momentum perpendicular
to the string axis. Thus the effective interaction area can be
enhanced at least ten times so that the suppression compared to
the first order phase transition scenario might be also improved.
\end{abstract}

\section{Introduction}

The application of the particle physics theory to the early stage of
the cosmic evolution provides various interesting ideas which can solve
some of the cosmological problems. One of the important examples is
the origin of the baryon asymmetry in the universe. The ratio of
baryon number density, $n_B$, to entropy density, $s$, is estimated as
\beq
\frac{n_B}{s} = 10^{-10\sim -9}
\eeq
by some observations\cite{KT,WSSOK}. Although there has been many
attempts and intensive study concerning the problem, we do not have
the final answer yet.

At present it seems to be probable that the baryon number is generated
at the scale of the unification of the electromagnetic and weak
interactions ($T\sim 10^2$ GeV), though other possibilities cannot be
neglected\cite{KT,CKN}. As pointed out by Sakharov\cite{Sak} there are
three necessary conditions for baryon asymmetry generation,
the existence of baryon number violating process, C and CP symmetry
violation, and the deviation from the thermal equilibrium.
In the electroweak unified theory, the first condition is satisfied
by electroweak anomaly. Sphaleron transitions violate the sum
of the baryon number and the lepton number\cite{spha}.
The second condition is also realized in the electroweak models.
The scenario which can solve the baryogenesis problem within the
Weinberg-Salam model is still proposed\cite{FS}. Note that, however,
quite a few people regard the degree of CP invariance violation
in the minimal standard model as too small to explain the amplitude
of the baryon asymmetry. Various types of model extension
are in progress in order to provide sufficient CP violation.

In the traditional electroweak baryogenesis, the out-of-equilibrium
condition is achieved by the expansion of the true vacuum bubbles.
However, such bubble nucleation occurs only when the order of the phase
transition equals one. It seems that the first order electroweak phase
transition is difficult to be established in the standard model.
Even if the non-standard extension of the theory enables the finite
temperature effective potential for the Higgs field to have the energy
barrier between the false vacuum and the true vacuum, thermal
fluctuations may prevent the bubble nucleation process\cite{SMY}.
The details of the electroweak phase transition is still in progress.

Recently one promising scenario of the electroweak baryogenesis has been
proposed\cite{BD}. Topological defects are another fruition of particle
cosmology\cite{Kib,topdef} and they can be an alternative to bubble
walls at the phase transition accompanied by supercooling. Electroweak
strings could be produced at the electroweak phase transition\cite{ews}.
These strings can be the source of the deviation from the thermal
equilibrium. Strictly speaking they are not topological defects so
that they should collapse into the trivial vacuum configuration.
During this process, the boundaries between the false vacuum and
the true vacuum move similarly to the walls of nucleated bubbles
in the first order phase transition. The interaction of these
boundaries with particles realizes the non-equilibrium condition.
The electroweak strings themselves have baryon number and can contribute
to the baryon asymmetry production\cite{ewsbn} or they can introduce
baryon number fluctuations through the interaction with the background
electromagnetic field\cite{Barr}. Moreover their effect on
the sphaleron transition rate has been discussed\cite{Soni}.

In this paper, we calculate the scattering cross section of quarks
from the electroweak strings and discuss the influence upon the
baryogenesis scenario. The remainder of the paper is organized as follows.
In the next section, the electroweak baryogenesis scenario by the collapses
of electroweak strings is briefly reviewed. The quark scattering cross
section from the electroweak strings is derived in the section 3.
The final section is devoted to discussion and conclusions.

\section{Baryogenesis by Electroweak Strings}

The electroweak strings are a kind of non-topological defect.
They are topologically unstable so that they have a monopole anti-monopole
pair at their ends or their configuration forms loops.
Although whether their configuration constitutes a local energy minimum
is under investigation\cite{stab}, it certainly satisfies the equations
of the minimal standard electroweak model. If they live out a time then
they can contribute to the baryogenesis as the sources of out-of-equilibrium
when the decay of the direction along the string axis which looks like
the monopole pair annihilation occurs.

In the standard model, the solutions of the Higgs field, $\Phi$, and
the spatial components of the Z-boson gauge field, ${\bf Z}$,
for a Z-string are written by
\bea
\Phi &=& \rho\left( r \right) \left(
\begin{array}{c}
0 \\ e^{i\theta}
\end{array}
\right)\ ,
\label{eq:phisol}\\
{\bf Z} &=& \frac{\alpha\left( r \right)}{r}
\left( -\sin \theta ,\ \cos \theta ,\ 0\right) \ , \label{eq:zsol}
\eea
where $r$ and $\theta$ are the radial component and the angular component
of the polar coordinate in the plane perpendicular to the string axis
and $\rho$ and $ \alpha$ are the Nielsen-Olesen solution for the U(1)
local string\cite{NO}. The equations which these functions obey are
\bea
& & \rho'' +\frac{\rho'}{r}-\frac{1}{r^2}\left( q_{\phi}\alpha
-1\right)^2 \rho-\lambda \left( \rho^2 -v^2\right) \rho=0\ , \\
& & \alpha'' -\frac{\alpha'}{r} -q_{\phi}\left( q_{\phi}\alpha -1\right)
\rho^2 =0\ ,
\eea
where the prime denotes a derivative by $r$,
$\lambda$ is the self-coupling constant for the Higgs field,
$v$ is the expectation value of the Higgs field at zero temperature,
and $q_{\phi}$ is the Z-charge of the Higgs field, {i. e.},
\beq
q_{\phi} =\frac{1}{2}\frac{e}{\sin \theta_w \cos \theta_w} \ ,
\eeq
where $\theta_w$ is the Weinberg angle. The other gauge fields
except ${\bf Z}$ have trivial configurations, {\it i. e.}, equal zero
everywhere.

The similar solution can be obtained for the W-string\cite{wews}.
However, the energy of the Z-string is lower than that of the W-string.
Hence we mainly consider the Z-string hereafter.

The baryon number density produced by the electroweak strings is estimated
to be that by the bubble nucleation times the suppression factor\cite{BDT}
\beq
\left. \frac{n_B}{s}\right\vert_{\rm string} \sim
\left. \frac{n_B}{s}\right\vert_{\rm bubble}
\times \frac{V_s}{V}\ , \label{eq:barasy}
\eeq
where $V_s$ is the volume swept out by the collapsing electroweak
strings and $V$ is the total volume of the universe. Since $V_s$ can
be regarded as the volume occupied by the strings just before
their collapse starts, it can be calculated as
\beq
V_s \simeq \delta_s^2 l_s \frac{V}{\xi_s^3}\ ,
\eeq
using the mean separation between strings, $\xi_s$, the averaged
length of the string, $l_s$, at the string formation epoch and
the width scale of the string, $\delta_s$. The length of the string
should be comparable to the mean separation, that is, $l_s \sim \xi_s$.
Then the suppression factor is represented by
\beq
\frac{V_s}{V}\sim \left( \frac{\delta_s}{\xi_s}\right)^2\ . \label{eq:supfac}
\eeq
The core radius of the electroweak string is almost the inverse of the
Higgs mass scale. If the number density of electroweak strings at the
formation epoch is described by the Kibble mechanism\cite{Kib}, then
the mean separation of them is equal to the correlation length of
the Higgs field, {\it i. e.}, the inverse Higgs mass scale. In this case,
$\xi_s \sim \delta_s$. Thus if the electroweak string distribution is
similar to that for the ordinary topological defect, the suppression factor
is almost of the order of unity.

In general, $\xi_s~\raisebox{-1ex}{$\stackrel{\textstyle >}{\sim}$}~
\delta_s$. For example, when $\xi_s$ is the particle horizon scale
at the electroweak phase transition, $V_s/V$ becomes $10^{-32}$.
The above estimation (\ref{eq:barasy}) is based on the calculation in
which the baryon number generation process is active only within
the string core. The magnification of the factor has been tried using
various particle physics models\cite{TDB}. Here we take into account
the amplification of the quark wave function near the electroweak string.

\section{Quark Interaction with Electroweak String}

In this section, we derive the scattering cross section of quarks from
the electroweak string, $\sigma_s$. The calculation method follows
that of fermion scattering\cite{fsfs}. In contrast to the calculations
in the references \cite{fsfs} where only the particles proceeding
perpendicular to the string were considered, we take into account
the momentum along the electroweak string axis, which will be proved to
be essential to the purpose of finding out the enhancement factor
of the baryogenesis by the electroweak strings. Note also that we are
interested in the total cross section, not in the helicity conserving
one or helicity flipping one. The result shows $\sigma_s$ is
not identical with a naive estimation, {\it i. e.}, the geometrical
cross section, $\delta_s^2$. It increases as the momentum of the particle
becomes nearly parallel with the string.

We express the quark spinor in the standard model as, the SU(2) quark
doublet : $\Psi_L^a$, the right-handed upper quark : $\psi_R^{+a}$, and the
right-handed lower quark : $\psi_R^{-a}$. Moreover each component of
$\Psi_L^a$ is written as $\psi_L^{+a}$ and $\psi_L^{-a}$, respectively.
The superscript, $a=1, 2, 3$, represents the family number. Since the
following formulae do not depend on the generation, we omit $a$ from now on.
The expressions of Yukawa coupling constants, $h^{+a}$ and $h^{-a}$, are also
simplified to $h^+$ and $h^-$. Then Dirac equations for quarks are written by
\bea
i\gamma^{\mu}D_{\mu}\Psi_L &=& h^+\tilde{\Phi}\psi_R^+ +h^-\Phi \psi_R^- \ ,\\
i\gamma^{\mu}D_{\mu}\psi_R^+ &=& h^+\tilde{\Phi}^{\dag}\Psi_L\ ,\\
i\gamma^{\mu}D_{\mu}\psi_R^- &=& h^-\Phi^{\dag}\Psi_L\ ,
\eea
where $\tilde{\Phi}\equiv i\sigma^2 \Phi^*$ and $D_{\mu}$ is
the covariant derivative.

The configurations of the Higgs field and the Z-boson gauge field are
arranged corresponding to the solutions of a static straight infinite
electroweak string, (\ref{eq:phisol}) and (\ref{eq:zsol}). Then in the
background where an electroweak string exists, Dirac equations for quarks
is modified to the formulae such as
\bea
& & i\gamma^{\mu}D_{\mu}\psi_L^+ -h^+\rho e^{-i\theta}\psi_R^+=0\ ,\\
& & i\gamma^{\mu}D_{\mu}\psi_L^- -h^-\rho e^{i\theta}\psi_R^-=0\ ,\\
& & i\gamma^{\mu}D_{\mu}\psi_R^+ -h^+\rho e^{i\theta}\psi_L^+=0\ ,\\
& & i\gamma^{\mu}D_{\mu}\psi_R^- -h^-\rho e^{-i\theta}\psi_L^-=0\ .
\eea

When the $\gamma$ matrix representation as
\beq
\gamma^0=\left(
\begin{array}{cc}
0 & 1 \\ 1 & 0
\end{array}
\right),\ \gamma^k=\left(
\begin{array}{cc}
0 & -\sigma^k \\ \sigma^k & 0
\end{array}
\right),
\eeq
is employed, the left-handed quarks have only two lower components
and the right-handed ones have two upper components which we write as
\beq
\psi_L^+=\left(
\begin{array}{c}
0 \\ w^+
\end{array}
\right),\
\psi_L^-=\left(
\begin{array}{c}
0 \\ w^-
\end{array}
\right),\
\psi_R^+=\left(
\begin{array}{c}
u^+ \\ 0
\end{array}
\right),\
\psi_R^-=\left(
\begin{array}{c}
u^- \\ 0
\end{array}
\right)\ .
\eeq
The equations are simplified to
\bea
& & \left( iD_0-i\sigma^kD_k\right) w^+ -h^+\rho e^{-i\theta}u^+=0\ , \\
& & \left( iD_0+i\sigma^kD_k\right) u^+ -h^+\rho e^{i\theta}w^+=0\ , \\
& & \left( iD_0-i\sigma^kD_k\right) w^- -h^-\rho e^{i\theta}u^-=0\ , \\
& & \left( iD_0+i\sigma^kD_k\right) u^- -h^-\rho e^{-i\theta}w^-=0\ .
\eea

For the moment we concentrate on the solutions for the lower quarks,
$w^-$ and $u^-$. The spinors can be decomposed to the eigen states
of the total angular momentum around the string axis as
\bea
u^- &=& \sum_{j=-\infty}^{+\infty}\left(
\begin{array}{c}
u_1^j(r) \\ iu_2^j(r)e^{i\theta}
\end{array}
\right) e^{ij\theta +ip_z z-i\omega t}\ , \\
w^- &=& \sum_{j=-\infty}^{+\infty}\left(
\begin{array}{c}
w_1^j(r) \\ iw_2^j(r)e^{i\theta}
\end{array}
\right) e^{i\left(j+1\right)\theta +ip_z z-i\omega t}\ ,
\eea
where $(r, \theta, z)$ are components of cylindrical coordinates whose
$z$-axis agrees with the string axis, $\omega$ is the total energy of
the quark and $p_z$ is the $z$-component of the quark momentum. Finally
the equations for each decomposed spinors are written by
\bea
& & \left( \frac{d}{dr}-\frac{j}{r}-2q_{\phi}q^-_R\frac{\alpha}{r} \right)
u_1^j +\left( \omega+p_z \right) u_2^j -h^-\rho w_2^j=0\ , \\
& & \left( \frac{d}{dr}+\frac{j+1}{r}+2q_{\phi}q^-_R\frac{\alpha}{r} \right)
u_2^j -\left( \omega-p_z \right) u_1^j +h^-\rho w_1^j=0\ , \\
& & \left( \frac{d}{dr}-\frac{j+1}{r}-2q_{\phi}q^-_L\frac{\alpha}{r} \right)
w_1^j -\left( \omega-p_z \right) w_2^j +h^-\rho u_2^j=0\ , \\
& & \left( \frac{d}{dr}+\frac{j+2}{r}+2q_{\phi}q^-_R\frac{\alpha}{r} \right)
w_2^j +\left( \omega+p_z \right) w_1^j -h^-\rho u_1^j=0\ ,
\eea
where $q^-_R$ and $q^-_L$ are the Z-charge of quarks :
\beq
q^-_R=\frac{1}{3}\sin^2 \theta_w\ ,\qquad
q^-_L=-\frac{1}{2}+\frac{1}{3}\sin^2 \theta_w \ .
\eeq

For simplicity we substitute a step-function like formula for $\rho(r)$ and
$\alpha(r)$ such as
\bea
\begin{array}{c}
\rho\left( r\right)=0 \\ \alpha\left( r\right)=0
\end{array}
\quad &;&
\qquad r\le \delta_s\ ,\\
\begin{array}{c}
\rho\left( r\right)=v \\ \alpha\left( r\right)=\frac{1}{q_{\phi}}
\end{array}
\quad &;&
\qquad r> \delta_s\ ,
\eea
that is, a trivial configuration inside the string core and the value
infinitely far from the string outside the core. Then the differential
equations take the form of the Bessel equations and the solutions can be
written by the Bessel functions.

Under the condition that the spinors must be regular at the origin, $r=0$,
the solutions inside the string is expressed by
\bea
u_1^j &=& c_j J_j\left( \sqrt{\omega^2-p_z^2}~r\right)\ , \label{eq:is1}\\
u_2^j &=& c_j D^+ J_{j+1}\left( \sqrt{\omega^2-p_z^2}~r\right)\ ,
\label{eq:is2}\\
w_1^j &=& d_j J_{j+1}\left( \sqrt{\omega^2-p_z^2}~r\right)\ ,
\label{eq:is3}\\
w_2^j &=& -d_j D^- J_{j+2}\left( \sqrt{\omega^2-p_z^2}~r \right)\ ,
\label{eq:is4}
\eea
where $c_j$ and $d_j$ are arbitrary constants and
\beq
D^+\equiv \sqrt{\frac{\omega-p_z}{\omega+p_z}}\ ,\qquad
D^-\equiv \sqrt{\frac{\omega+p_z}{\omega-p_z}}\ .
\eeq

On the other hand, the solutions outside the string core are written by
\bea
u_1^j &=& \left( a_j^1+b_j^1\right) J_{\nu}\left( R\right)
+\left( a_j^2+b_j^2\right) J_{-\nu}\left( R\right)\ , \label{eq:os1}\\
u_2^j &=& \left( a_j^1A^+-b_j^1A^-\right) J_{\nu+1}\left( R\right)
+\left( -a_j^2A^++b_j^2A^-\right) J_{-\nu-1}\left( R\right)\ ,
\label{eq:os2}\\
w_1^j &=& \left( a_j^1B^++b_j^1B^-\right) J_{\nu}
+\left( a_j^2B^++b_j^2B^-\right) J_{-\nu}\left( R\right)\ ,
\label{eq:os3}\\
w_2^j &=& \left( a_j^1A^+B^+-b_j^1A^-B^-\right) J_{\nu+1}\left( R\right)
+\left( -a_j^2A^+B^++b_j^2A^-B^-\right) J_{-\nu-1}\left(R\right)\ ,
\label{eq:os4}
\eea
where $a_j^1, a_2^j, b_1^j$ and $b_2^j$ are arbitrary constants and
\bea
A^+\equiv \sqrt{\frac{p-p_z}{p+p_z}}\ &,& \qquad
A^-\equiv \sqrt{\frac{p+p_z}{p-p_z}}\ , \\
B^+\equiv \frac{\omega-p}{m}\ &,& \qquad
B^-\equiv \frac{\omega+p}{m}\ ,
\eea
where $p=\sqrt{\omega^2-m^2}$ is the total momentum of the quark
and $m\equiv h^-v$ is its mass.
The argument of the above functions is defined as
\beq
R\equiv \sqrt{p^2-p_z^2}~r\ ,
\eeq
and the subscript is defined as
\beq
\nu \equiv j+2q^-_R=j+1+2q^-_L \ . \label{eq:nyu}
\eeq

Among six unknown coefficients four can be removed by the junction condition
of the internal solution (\ref{eq:is1}), (\ref{eq:is2}), (\ref{eq:is3}),
(\ref{eq:is4}) and the external solution (\ref{eq:os1}), (\ref{eq:os2}),
(\ref{eq:os3}), (\ref{eq:os4}). The other two should be determined by the
boundary condition. In order to calculate the scattering cross section,
we employ the incoming plain wave boundary condition at the infinity.
The plain wave of the $+$helicity is written by
\bea
u_1^j|_{r\to\infty} &=& \left( -i\right)^jJ_j\left( R\right) +f_j
\frac{e^{ip_{\perp}r}}{\sqrt{r}}+g_j\frac{e^{ip_{\perp}r}}{\sqrt{r}}\ , \\
u_2^j|_{r\to\infty} &=& i\left( -i\right)^jA^+J_{j+1}\left( R\right)
e^{i\theta} +f_j\frac{e^{ip_{\perp}r}}{\sqrt{r}}A^+e^{i\theta}-g_j
\frac{e^{ip_{\perp}r}}{\sqrt{r}}A^-e^{i\theta}\ , \\
w_1^j|_{r\to\infty} &=& \left( -i\right)^jB^+J_j\left( R\right) e^{i\theta}
+f_j\frac{e^{ip_{\perp}r}}{\sqrt{r}}B^+e^{i\theta}+g_j\frac{e^{ip_{\perp}r}}
{\sqrt{r}}B^-e^{i\theta}\ , \\
w_2^j|_{r\to\infty} &=& i\left( -i\right)^jB^+A^+J_{j+1}\left( R\right)
e^{2i\theta}+f_j\frac{e^{ip_{\perp}r}}{\sqrt{r}}B^+A^+e^{2i\theta}-g_j
\frac{e^{ip_{\perp}r}}{\sqrt{r}}B^-A^-e^{2i\theta}\ ,
\eea
and the $-$helicity plain wave is written by
\bea
u_1^j|_{r\to\infty} &=& \left( -i\right)^jJ_j\left( R\right) +f_j
\frac{e^{ip_{\perp}r}}{\sqrt{r}}+g_j\frac{e^{ip_{\perp}r}}{\sqrt{r}}\ , \\
u_2^j|_{r\to\infty} &=& \left( -i\right)^{j+1}A^-J_{j+1}\left( R\right)
e^{i\theta} +f_j\frac{e^{ip_{\perp}r}}{\sqrt{r}}A^+e^{i\theta}-g_j
\frac{e^{ip_{\perp}r}}{\sqrt{r}}A^-e^{i\theta}\ , \\
w_1^j|_{r\to\infty} &=& \left( -i\right)^jB^-J_j\left( R\right) e^{i\theta}
+f_j\frac{e^{ip_{\perp}r}}{\sqrt{r}}B^+e^{i\theta}+g_j\frac{e^{ip_{\perp}r}}
{\sqrt{r}}B^-e^{i\theta}\ , \\
w_2^j|_{r\to\infty} &=& \left( -i\right)^{j+1}B^-A^-J_{j+1}\left( R\right)
e^{2i\theta}+f_j\frac{e^{ip_{\perp}r}}{\sqrt{r}}B^+A^+e^{2i\theta}-g_j
\frac{e^{ip_{\perp}r}}{\sqrt{r}}B^-A^-e^{2i\theta}\ ,
\eea
where $p_{\perp}\equiv \sqrt{p^2-p_z^2}$ is the quark momentum perpendicular
to the string axis.

After a slight tedious calculation, the coefficients, $f_j$ and $g_j$,
can be obtained from which the scattering cross section per unit
string length, $d\sigma_s/dl$, can be deduced by
\beq
\frac{d\sigma_s}{dld\theta}=\sum_j \left| f_j\right|^2
+\sum_j \left| g_j\right|^2\ .
\eeq
Although the exact evaluation depends on $D^+, A^+$ and so on,
the important feature is that $d\sigma_s/dl$ is proportional to
the inverse of $p_{\perp}$ as
\beq
\frac{d\sigma_s}{dl} \propto \frac{1}{p_{\perp}}\ . \label{eq:sigmas}
\eeq
When $p_{\perp} \ll \delta_s^{-1}$ holds, the Z-charge of the quark appears as
\beq
\frac{d\sigma_s}{dl} \sim \frac{\sin^2\left( 2q^-_R\pi\right)}{p_{\perp}}\ .
\label{eq:sssin}
\eeq
Otherwise, that is, $p_{\perp} ~\raisebox{-1ex}
{$\stackrel{\textstyle >}{\sim}$}~\delta_s^{-1}$,
the factor $\sin^2\left( 2q^-_R\pi\right)$ disappears
which is not mentioned in the references \cite{fsfs}.

Now we return to the case of $w^+$ and $u^+$ .
The spinors can be decomposed as
\bea
u^+ &=& \sum_{j=-\infty}^{+\infty}\left(
\begin{array}{c}
\tilde{u}_1^j(r) \\ i\tilde{u}_2^j(r)e^{i\theta}
\end{array}
\right) e^{i\left(j+1\right)\theta +ip_z z-i\omega t}\ , \\
w^+ &=& \sum_{j=-\infty}^{+\infty}\left(
\begin{array}{c}
\tilde{w}_1^j(r) \\ i\tilde{w}_2^j(r)e^{i\theta}
\end{array}
\right) e^{ij\theta +ip_z z-i\omega t}\ ,
\eea
and the Z-charges of the upper quarks are
\beq
q^+_R=-\frac{2}{3}\sin^2 \theta_w\ ,\qquad
q^+_L=\frac{1}{2}-\frac{1}{3}\sin^2 \theta_w \ .
\eeq
Instead of the equation (\ref{eq:nyu}), the relation,
\beq
\nu' \equiv j+1+2q^+_R=j+2q^+_L \ ,
\eeq
holds so that the solutions for the upper quarks can be acquired
by the similar way as the lower ones and the scattering cross
section can be deduced. The result proves $d\sigma_s/dl$ has the same
characteristic as the equation (\ref{eq:sigmas}). The additional
factor when $p_{\perp} \ll \delta_s^{-1}$ is altered to
\beq
\frac{d\sigma_s}{dl} \sim \frac{\sin^2\left( 2q^+_L\pi\right)}{p_{\perp}}\ .
\label{eq:sssinp}
\eeq

In the standard model where $\sin^2 \theta_w \simeq 0.23$,
$2q^-_R \simeq 0.15$ and $2q^+_L \simeq 0.69$ so that the factors
$\sin^2(2q^-_R\pi)$ or $\sin^2(2q^+_L\pi)$ are not so significant.
Even if the non-standard value of $\sin^2 \theta_w$ is employed,
the essential feature that $\sigma_s$ is proportional to $1/p_{\perp}$
which can be seen in the equation (\ref{eq:sigmas}) is not altered.
Of course, when $\sin^2(2q^-_R\pi)\ll 1$ and $\sin^2(2q^+_L\pi)\ll 1$
hold, the interaction of the string with the quarks is suppressed
and the enhancement by small $p_{\perp}$ might be overcome.
Particularly when $\sin^2(2q^-_R\pi)$ and/or $\sin^2(2q^+_L\pi)$
equal exactly zero, the cross section has different dependence
from the equations (\ref{eq:sssin}) and (\ref{eq:sssinp}).

Such a situation can be realized within the standard model since
the change of $\sin^2 \theta_w$ has the same effect as the variation
of $q^-_R$ or $q^+_L$ in the calculation of the particle scatter.
The quark charge for the electroweak W-string is effectively
a half-integer, that is, $\sin(2q^-_R\pi)=0$ in the equation
(\ref{eq:sssin}). Then we have to use $N_{\nu}$ other than $J_{-\nu}$
in the equation (\ref{eq:os1}), (\ref{eq:os2}), (\ref{eq:os3}),
(\ref{eq:os4}) since $J_{\nu}$ and $J_{-\nu}$ are linearly dependent
when $\nu$ is an integer. For the reason that an integer-order Neumann
function has an logarithmic dependence, an extra factor comes out as
\beq
\frac{d\sigma_s}{dl} \sim \frac{1}{p_{\perp}}\frac{1}
{\ln^2 \frac{p_{\perp}\delta_s}{2}}\ ,
\eeq
when $p \gg m$ besides $p_{\perp} \ll \delta_s^{-1}$. This factor is
neither important since $\left( \ln \frac{p_{\perp}\delta_s}{2}\right)^{-2}$
becomes very small when $p_{\perp} \ll \delta_s^{-1}$. We are interested in
the cases that $\sigma_s$ should be enhanced.

Finally we take a brief look at the situation when the quark proceeds
in parallel with the string axis. When $\omega=\sqrt{p_z^2+m^2}$ as
usual, the internal solution is written by the Bessel functions
similarly to the formulae (\ref{eq:is1}), (\ref{eq:is2}),
(\ref{eq:is3}), (\ref{eq:is4}) and the solution outside the string
core damps in the power law manner as $r^{\nu}$, $r^{\nu+1}$,
$r^{-\nu}$, $r^{-\nu-1}$. Under the condition that $\omega=p_z$, the
inner solution is expressed by the power law functions or constants 
and the outer one damps exponentially as $e^{-mr}$.
It is natural that the smaller the momentum perpendicular to the string axis
the larger the amplitude of the wave function of the particle inside the string
since $p_{\perp}$ supplies the energy which is converted to the mass, $m$,
outside the string.

\section{Conclusions}

In the previous section, we have calculated the scattering cross section of
the quarks from the electroweak strings. The result can be summarized as
\beq
\frac{d\sigma_s}{dl} \sim \frac{1}{p_{\perp}}\ .
\eeq

The length along the string where the baryogenesis occurs sufficiently
can be estimated to be at least $\sim \delta_s$ since only the region
where the boundary between the false vacuum and the true vacuum is
in question. Hence the quark scattering cross section should be no less 
than
\beq
\sigma_s \sim \frac{\delta_s}{p_{\perp}}\ .
\eeq
As the next step we calculate the averaged value of $1/p_{\perp}$.
If the homogeneous probability distribution of $p_{\perp}$ is assumed then
\beq
\left< \frac{1}{p_{\perp}}\right> =\frac{1}{p}\int_{\Lambda}^p
\frac{1}{p_{\perp}}dp_{\perp}=\frac{1}{p}\ln\left(
\frac{p}{\Lambda}\right)\ ,
\eeq
where $\Lambda$ is the infrared cutoff scale. Since the momentum of the quark
should be the scale of the temperature at the electroweak phase transition,
{\it i. e.}, the scale of the Higgs mass $\sim \delta_s^{-1}$,
\beq
\left< \frac{1}{p_{\perp}}\right> \sim \delta_s
\ln\left( \frac{\Lambda^{-1}}{\delta_s}\right)\ .
\eeq
Thus the enhance factor of $\sigma_s$ relative to the geometrical cross
section, $\delta_s^2$, is
\beq
\ln\left( \frac{\Lambda^{-1}}{\delta_s}\right) \sim 40\ ,
\eeq
when $\Lambda$ is taken to be the inverse of the horizon scale at the
electroweak symmetry restoration epoch, $t\sim 10^{14}$ GeV$^{-1}$.
This is not a radical amplification. We, however, treat the quarks
from all the direction equivalently. The most effective particle
for the baryon number production is one which is running against
the string end vertically. The damping of the wave function
in the region outside the string core suggests that $\sigma_s$
can be expected to be much larger.

Therefore the effective increase of the interaction area between strings
and particles may partially cancel the smallness of the suppression factor
(\ref{eq:supfac}). Such an enhancement is probable since the mass of the
quarks vanishes in the string core where the electroweak symmetry
is restored so that they prefer to reside inside the string. Even if
the Kibble mechanism is valid and the suppression factor is not so small,
the increase of the produced baryon asymmetry estimation generally
means that the needed CP violation strength can be lower and
the constraints on the model should be relaxed .

The estimation in this paper is only rough one. The precise value of the
interaction cross section should depend on the basic process of the baryon
generation. Further investigation might reveal the usefulness of
the baryogenesis by the electroweak strings.

\section*{Acknowledgments}

The author is grateful to Professor J. Yokoyama for his continuous
encouragement. This work was partially supported by the Japanese Grant
in Aid for Science Research Fund of the Ministry of Education, Science
and Culture (No. 5110).


\begin{thebibliography}{30}
\bibitem{KT}
E.\ W.\ Kolb and M.\ S.\ Turner, {\it The Early Universe},
(Addison-Wesley Publishing Company, Redwood City, 1990) p.157
and references therein.
\bibitem{WSSOK}
T.\ P.\ Walker, G.\ Steigman, D.\ N.\ Schramm, K.\ A.\ Olive
and H.-S.\ Kang, Astrophys.\ J.\ {\bf 376}, 51\ (1991).
\bibitem{CKN}
A.\ G.\ Cohen, D.\ B.\ Kaplan and A.\ E.\ Nelson,
Ann.\ Rev.\ Nucl.\  Part.\ Sci.\ {\bf 43}, 27\ (1993).
\bibitem{Sak}
A.\ D.\ Sakharov, JETP\ Lett.\ {\bf 5}, 24\ (1967).
\bibitem{spha}
G.\ 't\ Hooft, Phys.\ Rev.\ Lett.\ {\bf 37}, 8\ (1976).\\
M.\ S.\ Manton, Phys.\ Rev.\ {\bf D28}, 2019\ (1983).
\bibitem{FS}
G.\ R.\ Farrar and M.\ Shaposhnikov, Phys.\ Rev.\ {\bf D50}, 774\ (1994).
\bibitem{SMY}
T.\ Shiromizu, M.\ Morikawa and J.\ Yokoyama,
Prog.\ Theor.\ Phys.\ {\bf 94}, 795\ (1995).
\bibitem{BD}
R.\ H.\ Brandenberger and A.-C.\ Davis,
Phys.\ Lett.\ {\bf B308}, 79\ (1993).
\bibitem{Kib}
T.\ W.\ B.\ Kibble, J.\ Phys.\ {\bf A9}, 1387\ (1976).
\bibitem{topdef}
T.\ W.\ B.\ Kibble, Phys.\ Rep.\ {\bf 67}, 183\ (1980).\\
A.\ Vilenkin, Phys.\ Rep.\ {\bf 121}, 263\ (1985).\\
A.\ Vilenkin and E.\ P.\ S.\ Shellard, {\it Cosmic Strings and Other
Topological Defects}, (Cambridge University Press, New York, 1994).
\bibitem{ews}
Y.\ Nambu, Nucl.\ Phys.\ {\bf B130}, 505\ (1977).\\
V.\ Soni, Phys.\ Lett.\ {\bf B93}, 101\ (1980).\\
A.\ Sugamoto, Phys.\ Lett.\ {\bf B127}, 75\ (1983).\\
T.\ Vachaspati, Phys.\ Rev.\ Lett.\ {\bf 68}, 1977\ (1992).\\
T.\ Vachaspati, Nucl.\ Phys.\ {\bf B397}, 648\ (1993).\\
G.\ Dvali and G.\ Senjanovi\'{c}, Phys.\ Rev.\ Lett.\ {\bf 71},
2376\ (1993).\\
G.\ Dvali and G.\ Senjanovi\'{c}, Phys.\ Lett.\ {\bf B331}, 63\ (1994).
\bibitem{ewsbn}
T.\ Vachaspati, Phys.\ Rev.\ Lett.\ {\bf 73}, 373\ (1994).\\
M.\ Sato and S.\ Yahikozawa, Nucl.\ Phys.\ {\bf 436}, 100\ (1995).
\bibitem{Barr}
M.\ Barriola, Phys.\ Rev.\ {\bf D51}, R300\ (1995).
\bibitem{Soni}
V.\ Soni, preprint NPL-TH-94-1, hep-ph/9505386\ (1995).
\bibitem{stab}
M.\ James, L.\ Perivolaropoulos and T.\ Vachaspati,
Phys.\ Rev.\ {\bf D46}, R5232\ (1992).\\
W.\ B.\ Perkins, Phys.\ Rev.\ {\bf D47}, R5224\ (1993).\\
M.\ James, L.\ Perivolaropoulos and T.\ Vachaspati,
Nucl.\ Phys.\ {\bf B395}, 534\ (1993).\\
T.\ Vachaspati and R.\ Watkins, Phys.\ Lett.\ {\bf B318}, 163\ (1993).\\
A.\ Ach\'{u}carro, R.\ Gregory, J.\ A.\ Harvey and K.\ Kuijken,
Phys.\ Rev.\ Lett.\ {\bf 72}, 3646\ (1994).\\
M.\ A.\ Earnshaw and W.\ B.\ Perkins, Phys.\ Lett.\ {\bf B328}, 337\ (1994).\\
J.\ M.\ Moreno, D.\ H.\ Oaknin and M.\ Quir\'{o}s, Phys.\ Lett.\ {\bf B347},
332\ (1995).\\
G.\ Bimonte and G.\ Lozano, Phys.\ Lett.\ {\bf B348}, 457\ (1995).\\
S.\ G.\ Naculich, Phys.\ Rev.\ Lett.\ {\bf 75}, 998\ (1995).\\
L.\ Masperi and A.\ M\'{e}gevand, preprint hep-ph/9410211\ (1996)
to be published in Zeitschrift fur Physik C.
\bibitem{NO}
H.\ B.\ Nielsen and P.\ Olesen, Nucl.\ Phys.\ {\bf B61}, 45\ (1973). 
\bibitem{wews}
T.\ Vachaspati and M.\ Barriola, Phys.\ Rev.\ Lett.\ {\bf 69},
1867\ (1992).\\
M.\ Barriola, T.\ Vachaspati and M.\ Bucher, Phys.\ Rev.\ {\bf D50},
2819\ (1994).
\bibitem{BDT}
R.\ H.\ Brandenberger, A.-C.\ Davis and M.\ Trodden,
Phys.\ Lett.\ {\bf B335}, 123\ (1994).
\bibitem{TDB}
M.\ Trodden, A.-C.\ Davis and R.\ H.\ Brandenberger,
Phys.\ Lett.\ {\bf B349}, 131\ (1995).
\bibitem{fsfs}
N.\ Ganoulis, Phys.\ Lett.\ {\bf B298}, 63\ (1993).\\
A.\ C.\ Davis, A.\ P.\ Martin and N.\ Ganoulis, Nucl.\ Phys.\ {\bf B419}
323\ (1994).\\
H.-K.\ Lo, Phys.\ Rev.\ {\bf D51}, 802\ (1995).
\end{thebibliography}
\end{document}